\documentclass{article}

\usepackage{url}
\usepackage{todonotes}
\usepackage{xspace}
\usepackage{numprint}
 \usepackage{subcaption}
\usepackage{algorithm,algorithmic}

\author{Matthias Gall\'e}

\title{Multi-view pattern matching}

\begin{document}

\maketitle

\begin{abstract}
We introduce the \textit{multi-view pattern matching} problem, where a text can have multiple views.
Each view is a string of the same size and drawn from disjoint alphabets.
The pattern is drawn from the union of all alphabets.

The algorithm we present is an extension of the Horspool algorithm, and in our experiments on synthetic data it shows an $3 \times$ improvement over the naive baseline.
\end{abstract}

\section{Motivation}

Suppose we are given a text, together with its Parts-of-Speech (POS) annotated sequence, and want to retrieve all occurrences of \textit{Carlson invented NN}.
Another example is the retrieval of records for a collection of trivia for instance: supposing an additional second layer of annotation where each word is annotated if it is a superlative (\textit{SUP}) or not, and we are interested in the occurrences of \textit{NNS is the SUP NNS of France} (Mt-Blanc, highest, mountain; Bettencourt, wealthiest, person, Saint-Cirq-Lapopie, prettiest, village; etc).

Although pattern matching is one of the most studied problems in computer science, currently there are no methods to solve this problem directly. 
Regular expression in particular are not suited for this kind of problem because the variety lies in the sequence, not in the pattern.

We call this the \textit{multi-view pattern matching} problem, because we consider that there are $k$ different views over the sequence $S$, and the pattern can contain any combination of these views.
In this paper we present an efficient method that address the problem of finding all occurrences of $p$ in such a multi-view sequence $S$ by pre-processing $p$.

\subsection{Problem Definition}

We are given $k$ sequences $S=[s_1, \dots, s_k]$, all of size $n$ and over pair-wise disjoint alphabets ($s_i \in \Sigma_i^*$; $\Sigma_i \cap \Sigma_j = \emptyset, \forall i \neq j$).
We denote by $t(c)$ the type of character $c$: $t(c) = k$ iff $c \in \Sigma_k$.

A pattern $\displaystyle p \in \left(\bigcup_{i=1}^k \Sigma_i\right)^*$ occurs at position $i$ of $S$ if $p[j] = s_{t(p[j])}[i+j]$, $\forall 1 \leq j \leq |p|$.
The problem we then want to solve is to find all occurrences of $p$ in $S$.

%In general pattern matching algorithms differ depending if the sequence $S$ is considered stable and several different patterns will be searched on it (in which it may be more efficient to index the sequence); or rather the pattern $p$ is fixed and it will be queried over different sequences, or the $p$ is significantly shorter than $S$ (in which case it is convenient to pre-process the pattern).

%\section{Indexing $S$}

\section{Related Work}
In broad terms, pattern matching algorithms divide into two classes: if the sequence $S$ is considered stable and several different patterns will be searched on it it may be more efficient to index the sequence $S$.
But if the pattern $p$ is fixed and it will be queried over different sequences, or $p$ is significantly shorter than $S$ it is convenient to pre-process the pattern instead.
Here we focus on this second case.

There are several different exact string matching algorithms, and many variations on each of them\footnote{these lecture notes for instance describe 28 such algorithms: \url{http://www-igm.univ-mlv.fr/~lecroq/string/index.html}}.
The two most famous ones are probably the Knuth-Morris-Pratt~\cite{KMP} and the Boyer-Moore~\cite{BM1977} ones.
Both improve over a brute-force algorithm by taking advantage of the information of a mismatch, shifting over parts of $S$ on which one can be certain that a match will not occur.

To see the intuition behind, consider such a brute-force algorithm, that tries to match $p$ at each position $i$ of $S$. 
If for instance $p=\textit{abc}$ and $S=\textit{abdabc}$, the first alignment would fail because $S$ has a $d$ in its third position, compared to an $c$ for $p$. 
The pattern would then be shifted by one, and $bda$ (substring starting at position 2 of $S$) would be compared to $p$.
However, at this stage we already know that $S$ does not contain any $a$ (the first letter of the pattern) in its three first position, so aligning $p$ to any of these positions fail for sure.
KMP is based upon this idea, creating a \textit{failure table} at pre-processing time that gives for each position $j$ of $p$ the position to look for a new match if the current one failed.
So, when \textit{ababc} is matched against \textit{abababc}, the first alignment (at position $1$ of $S$) will fail at position $5$ of $p$ ($c \neq a$), but the failure table will indicate that the next alignment should start at position $3$ of $S$, because it observed there \textit{ab} which coincides with the start of $p$.

\begin{figure}
	\centering
	\includegraphics[width=200pt]{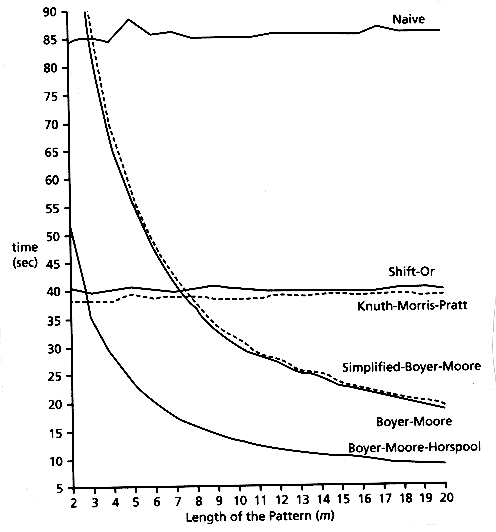}
	\caption{Comparison of exact string matching algorithms (extracted from~\cite{Frakes1992}). This uses random generated text, but the figure for English text in the book is almost equivalent.}
	\label{fig:comparison} 
\end{figure}

As can be seen from that example, this algorithm takes advantage of repetitions inside $p$ (in particular repetitions of any prefix of $p$). 
This becomes even more explicit in the case of the Boyer-Moore algorithm.
The original one had two different tables, and the second one focuses precisely on handling such repetitions of suffixes (because BM compares from right to left).
However such repetitive patterns are unusual in practice.
This is precisely the insight of Horspol~\cite{Horspool1980} who shows how a simple version of BM (dropping the second rule, and further simplifying the first one) provides sometimes results even better than BM because it shortens the pre-processing time.
Fig.~\ref{fig:comparison} extracted from~\cite{Frakes1992} compares different exact simple string pattern matching (Boyer-Moore-Horspool there correspond to~\cite{Horspool1980}).

\smallskip

Much work has been done to support inexact (or fuzzy) pattern matching, including cases when $p$ includes wildcards (fixed or extensible), multi-character symbols and regular expressions.
However, for the precise case where the variations are in $S$ under the form we are considering here we were not able to find any work addressing it.

\section{Multi-View Horspool}

We decided to focus on the Horspool algorithm and extending it for the multi-view case for two main reasons:

\begin{enumerate}
	\item Its good performance on benchmarks (see again Fig.~\ref{fig:comparison})
	\item The improvement of other algorithms with respect to it lie in the fact that they avoid comparisons due to repetitions inside the pattern. However in our case of multi-view patterns, this is not so straigthforward as the following examples shows.
\end{enumerate}

$p=\textit{AaAab}$ and $S=$

\[ \begin{array}{ccccccc}
1 & 2 & 3 & 4 & 5 & 6 & \dots\\
\hline
b & a & a & a & a & b & \dots\\
A & A & A & A & B & B & \dots\\
\end{array} \]

Aligning $p$ at position 1 fails due to the last character.
However, if we would take advantage of the repetition of \textit{Aa} and shift $p$ directly to the position $3$, we would be losing the match that occures at position $2$.
The fact that symbols from one alphabet may mask symbols from another makes such an approach much more cumbersome.
They act as a kind of wildcard in the pre-processing (but not during the matching), and existing extensions of pattern matching to such cases are very sensitive to the number of such wildcards (in our case, any character acts as wildcard for the other alphabtes)\footnote{see for instance the problems in these lecture notes from Jeff Erickson's class at Urbana-Champaign \url{http://www.cs.uiuc.edu/class/fa06/cs473/lectures/18-kmp.pdf}}.

\bigskip

Like in Horspool's algorithm, the pre-processing we propose consists in filling a table $\beta$ (for \textit{bad character}, the name of the first rule of BM) that gives for each symbol the offset from the end of its latest occurrence.
For the previous example this would be $\beta(a) = 1, \beta(A) = 2, \beta(b) = 5, \beta(B)=1$ (the last occurrence of $A$ is at position $4 = 6-\mathbf{2}$).
Note that for the characters occurring in the last position it gives the offset to the second-to-last occurrence (the reason for this will become clear later on), or $m=|p|$ if it does not occur.
In general 

\begin{equation}
\beta(c) = \min\{ i \; | \;  i = m \vee (1 \leq i < m \wedge p_{t(c)}[m-i] = c )\}
\label{eq:preprocess}
\end{equation}

The algorithm we propose is then as follows (see also Alg.~\ref{alg:mvhorspool}).
For each alignment, we first compare the last character and, if it coincides, we then continue comparing the remaining ones.
This is equivalent to the classical Horspool algorithm. 
The novelty lies in the shift, where we consider the lowest value of all the offsets of symbols occurring at the current position, taken over all sequences $s_k$.

\begin{algorithm}{mv-horspool$(S,p,t)$}
	\caption{}
	\label{alg:mvhorspool}

	\begin{algorithmic}[1]
		\REQUIRE{sequence $S=[s_1, \dots, s_k$, pattern $p$, and implicit type function $t$}
		\ENSURE{all matches of $p$ in $S$}
			\STATE pre-process $p$ to produce $\beta$ holding Eq.~\ref{eq:preprocess}
			\STATE $j := 0$
			\STATE $n, m := |S|, |p|$
			\WHILE{$j \leq n-m$}
				\STATE $c := S_{t(p[m])}[j + m]$
				\IF{$p[m] = c \wedge p[1:m-1] \textit{ matches } S[j:j+m-1]$}
						\STATE \textbf{output} $j$
				\ENDIF
				\STATE $j := j + \min_k \beta(S_k[j + m]) $
			\ENDWHILE
	\end{algorithmic}
\end{algorithm}

Consider its execution on the following example:

$p=\textit{BAbB}$ (and therefore $\beta(B)=3, \beta(A)=2, \beta(b)=1$ and $m=4$ for all other symbols); and $S=$

\[ \begin{array}{cccccccc}
1 & 2 & 3 & 4 & 5 & 6 7 & 8 \\
\hline
c & a & b & b & a & a & b & c  \\
B & A & B & A & B & A & C & B\\ 
\end{array} \]

In the first alignment, the last character already does not match:

\[ \begin{array}{cccccccc}
1 & 2 & 3 & 4 & 5 & 6 & 7 & 8 \\
\hline
c & a & b & b & a & a & b & c  \\
B & A & B & {\color{red}A} & B & A & C & B\\ 
\hline
B & A & b & {\color{red}B} \\
\end{array} \]

For the shift we consider the lowest value of $\beta(b)$ and $\beta(A)$ which is 1 and start the second alignment:

\[ \begin{array}{cccccccc}
1 & 2 & 3 & 4 & 5 & 6 & 7 & 8 \\
\hline
c & a & b & b & a & a & b & c  \\
B & {\color{red}A} & B & A & {\color{green}B} & A & C & B\\ 
\hline
& {\color{red}B} & A & b & {\color{green}B} \\
\end{array} \]

The last character matches, but not the remaining one (here we suppose that we compare from the beginning to the end).
$p$ is then shifted according to  $\min(\beta(s_1[5],s_2[5])$, which is $3$ (as $\beta(a)=m=4$).
This finally results in a match:

\[ \begin{array}{cccccccc}
1 & 2 & 3 & 4 & 5 & 6 & 7 & 8 \\
\hline
c & a & b & b & a & a & {\color{green}b} & c  \\
B & A & B & A & {\color{green}B} & {\color{green}A} & C & {\color{green}B}\\ 
\hline
& & & & {\color{green}B} & {\color{green}A} & {\color{green}b} & {\color{green}B} \\
\end{array} \]

\section{Results}

Because Alg.~\ref{alg:mvhorspool} has to compare the values at the current position of all $k$ sequences, it is not clear that this should be faster than the naive algorithm, which is independent of $k$.

\begin{figure}
	\centering
	\includegraphics[width=200pt]{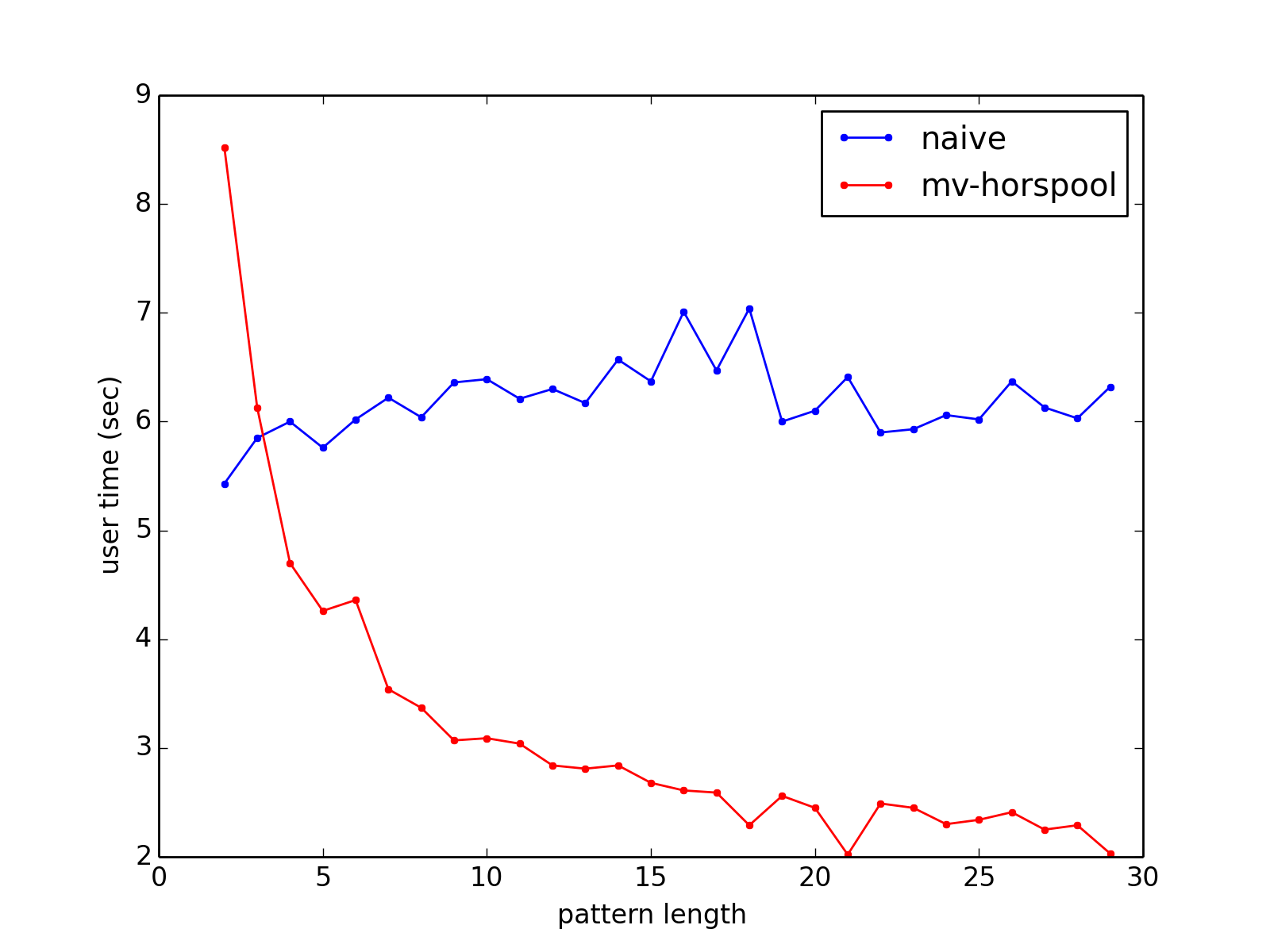}
	\caption{Total user time for \numprint{10000} instances of multi-view pattern matching.}
	\label{fig:mvcomparison} 
\end{figure}

We therefore compare the performance over randomly generated text, using the following setting: $k=3, n=\numprint{100000}, |\Sigma_i|=10$ and varying $m$ from $2$ to $30$.
We implemented both algorithms in $C$, and measured user time over \numprint{10000} generated examples (same sequences for each algorithm, generated uniformly and i.i.d).
The results can be appreciated in Fig.~\ref{fig:mvcomparison}

While the improvement is not so impressive compared to the standard case (with has improvement of up to x8, see Fig.~\ref{fig:comparison}), the performance improvement is still of a factor of 3 times better than the naive algorithm.

\bibliographystyle{plain}
\bibliography{biblio_db}

\end{document}